\documentstyle[12pt,aps]{revtex}

\newcommand{\be}{\begin{equation}}
\newcommand{\bea}{\begin{eqnarray}}
\newcommand{\ee}{\end{equation}}
\newcommand{\eea}{\end{eqnarray}}

\begin{document}
\topmargin -1cm
\oddsidemargin=0.25cm\evensidemargin=0.25cm
\renewcommand{\thefootnote}{\fnsymbol{footnote}}
\thispagestyle{empty}
\begin{center}
{\large\bf  Reparametrization Invariance and the Schr\"odinger Equation }
\vspace{0.5cm} \\
{\bf V.I. Tkach}\footnote{E-mail: vladimir@ifug1.ugto.mx}
\vspace{0.5cm}\\
{\it Instituto de F\'{\i}sica de la Universidad de Guanajuato,\\
Apartado Postal E-143, C.P. 37150, Leon Gto. Mexico}
\vspace{0.2cm}\\
{\bf A.I. Pashnev}\footnote{E-mail: pashnev@thsun1.jinr.ru},
and {\bf J.J. Rosales}\footnote{E-mail: rosales@thsun1.jinr.ru}
\vspace{0.2cm} \\
{\it JINR--Bogoliubov Laboratory of Theoretical Physics, \\
141980 Dubna, Moscow Region, Russia}
\vspace{1cm}\\

{\bf Abstract}
\end{center}
In the present work we consider a time-dependent Schr\"odinger
equation for systems invariant under the reparametrization of
time. We develop the two-stage procedure of construction such
systems from a given initial ones, which
 is not invariant under the time reparametrization. One of the first-class
constraints of the systems in such description becomes the time-dependent
Schr\"odinger equation. The procedure is applicable in the supersymmetric
theories as well.  The $n=2$ supersymmetric
quantum mechanics is coupled to world-line supergravity, and the local
supersymmetric action is constructed leading to the square root representation
of the time-dependent Schr\"odinger equation.

\vspace{1cm}

\vspace{0.5cm}

\newpage
\section{Introduction}

Time plays a central and peculiar role in Hamiltonian quantum mechanics.
In the standard non-relativistic quantum mechanics one can describe the
motion of a system by using the canonical variables which are only
functions of time. The scalar product specifies a direct probability of
observation at one instant of time \cite{1}. Time is the sole observable
assumed to have a direct physical significance, but it is not a dynamical
variable itself. It is an absolute parameter differently treated from the
other coordinates, which turn out to be operators and observables in quantum
mechanics.

In the cases of non-relativistic and relativistic point particles
mechanics generally covariant systems may be obtained by promoting
the time $t$ to a dynamical variable \cite{1,2,3,4,5,6,7,8}. The
idea behind this transformation is to treat symmetrically the time
and dynamical variables. This is achieved by taking the time $t$
as a function of an arbitrary parameter $\tau$ (label time) in
Dirac's approach \cite{2}. The arbitrariness of the label time
$\tau$ is reflected in the invariance  of the action under the
$\tau$ reparametrization.

In this work we give the two-stage procedure for constructing
generally covariant systems. Using additional gauge variables we
rewrite the original action of the system in the reparametrization
invariant form \cite{2,3}. The structure of the reparametrization
transformations leads to zero Hamiltonian (first-class constraint)
associated to the original action \cite{3,6}. At the quantum
theory this constraint imposes condition on the  vector states,
which becomes time-independent Schr\"odinger equation \cite{3,8}.
After that we consider an additional action invariant under
reparametrization, which does not change the equations of motion
of the original action, but modifies only the first-class
constraint, which becomes now the time-dependent Schr\"odinger
equation \cite{3,5}. In the case of different versions of
supersymmetric quantum mechanics \cite{9,10,11} such a procedure
finds its application, when the transformations of
reparametrization belong to a wider group of local transformations
arising from the construction of the generally covariant systems.
In this case, the set of auxiliary gauge variables are components
of the world-line supergravity multiplet \cite{19}.

 Here we construct a local
supersymmetric action for $n=2$, $d=1$ supersymmetric quantum
mechanics, in which the first-class constraint becomes
time-independent Schr\"odinger equation, supercharges and the
fermion number operator. It is well known, that in the case of
supersymmetric quantum mechanics there is a square root
representation for the vector states of the original Hamiltonian,
a state with zero energy \cite{9,10,11,19}. It will be shown, that
there exists an additional supersymmetric invariant action, which
permits the generalization of the above local supersymmetric
quantum theory. Hence, we have the square root representation of
the Schr\"odinger operator.

The plan of this work is as follows: in section 2, applying the canonical
quantization procedure to reparametrization invariant action, we obtain
the time-dependent Schr\"odinger equation. In section 3 the same procedure
is applied to relativistic case. The extension to supersymmetric
model is performed in section 4. Finally, section 5 is devoted to final
remarks.

\setcounter{equation}0\section{Non-relativistic parametrized
particle dynamics} In this section the central idea is illustrated
with the aid of a simple model of parametrized dynamics.

We start by considering the theory of a non-relativistic particle moving
in the three dimensional space with dynamical variables $x_i$
$ (i= 1, 2, 3)$ and with $t$ denoting the ordinary physical time parameter.
The action for this simplest model may be written as
\be
S_0 = \int\left\{ \frac{1}{2}m \dot x_i^2(t) - V(x_i)\right \} dt,
\label{1}
\ee
where $m$ is the mass of the particle, $\dot x_i = \frac{dx_i}{dt}$ is its
velocity and $V(x_i)$ is the potential. The action (\ref{1}) is invariant
under the global translation of time
\be
t^{\prime} \to t + c, \qquad c= \mbox{constant}.
\label{2}
\ee
We see, that the Lagrangian is non-degenerate in the sense that the
relation between momentum and velocity is one to one
\be
p_i = \frac{\partial L}{\partial {\dot x^i}} = m\dot x_i.
\label{3}
\ee
The Hamiltonian for this model has the form
\be
H_0 = \frac{p^2_i}{2m} + V(x_i).
\label{4}
\ee
In the action (\ref{1}) time $t$ is an absolute parameter, differently
treated from the other coordinates which turn out to be operators and
observables in quantum mechanics. On the other hand, it is well known, that
in non-relativistic point particle mechanics generally covariant systems
may be obtained by promoting the time $t$ to a dynamical variable \cite{2,3}.
The same procedure has been applied to relativistic particle case
\cite{6,7}.

So, we will rewrite the action (\ref{1}) in the parametrized form
\be
\tilde S = \int \left \{ \frac{m \dot x_i^2(\tau)}{2 N(\tau)} -
N(\tau) V[x_i(\tau)]\right \} d\tau, \label{5} \ee where the dot
denotes derivative with respect to the parameter $\tau$. $N(\tau)$
is the so called ``lapse function" and relates the physical time
$t$ to the arbitrary parameter $\tau$ through $dt = N(\tau)d\tau$.
This canonical variable is a pure gauge variable and it is not
dynamical. $N(\tau)$ in (\ref{5}) defines the scale on which the
time is measured, and in the ``gauge" $N(\tau) = 1$ the time
parameter $\tau$ is identified as the ``classical" time $t$ and
(\ref{5}) becomes (\ref{1}). On the other hand, $N(\tau)$ can be
viewed as one dimensional gravity field, then the action (\ref{5})
describes the interaction between ``matter"  $x_i(\tau)$ and the
gravity field $N(\tau)$ \cite{12}. The action (\ref{5}) is
invariant under the local conformal time transformation
\be
\tau^{\prime} = \tau + a(\tau),
\label{6}
\ee
if $N(\tau)$ and $x(\tau)$ transform as
\be
\delta N(\tau) = (aN)^. \qquad \qquad \delta x_i(\tau) = a \dot
x_i(\tau). \label{7} \ee This is because $\delta \tilde S =
\int\frac{d}{d\tau}(a\tilde L) d\tau $ is a total derivative with
the Lagrangian $\tilde L = \frac{m {\dot x_i}^2}{2N} - NV(x_i)$.

Varying the action (\ref{5}) with respect to $x(\tau)$ and
$N(\tau)$ one obtains the classical equations of motion for $x(\tau)$ and
the constraint, respectively. The constraint generates the local
reparametrization of $x(\tau)$ and $N(\tau)$.

Now we consider the Hamiltonian analysis of this simple constrained system.
We define the canonical momentum $p^i$ conjugate to the dynamical variable
$x_i$ as
\be
p^i = \frac{\partial \tilde L}{\partial \dot x_i} = \frac{m}{N} \dot x^i,
\label{8}
\ee
and the classical Poisson brackets between $x_i$ and $p^j$ by
\be
\lbrace x_i, p^j \rbrace = \delta_i^j.
\label{9}
\ee
The momentum conjugate to $N(\tau)$ is
\be
P_N = \frac{\partial \tilde L}{\partial{\dot N}} = 0,
\label{10}
\ee
this equation merely constrains the variable $N(t)$ (primary constraint). The
canonical Hamiltonian can be calculated in the usual way, it has the form
$\tilde H_c = NH_0$, and the total Hamiltonian is
\be
\tilde H_T = NH_0 + u_N P_N,
\label{11}
\ee
where $u_N$ is the Lagrange multiplier associated to the constraint
$P_N = 0$ in (\ref{10}) and $H_0$ is the Hamiltonian of the system
defined in (\ref{4}). The canonical evolution of the constraint $P_N$ is
given by the Poisson bracket with the total Hamiltonian. Thus, we have
\be
\dot P_N = \lbrace P_N, \tilde{H}_c + u_NP_N \rbrace = - H_0 = 0,
\label{12}
\ee
leading to the secondary constraint, which by definition is of the
first-class constraint \cite{5}. In the quantum theory the first-class
constraint associated with the invariant action (\ref{5}) under the
transformations of reparametrization (\ref{6}) becomes condition on the
wave function $\psi$. So that any physical state must obey the following
quantum constraint
\be
H_0(\hat{p}^i, x_i) \psi(x_l) = 0,
\label{13}
\ee
which is nothing but the time-independent Schr\"odinger equation.

Now we have to stress, that the physical meaning of the action
(\ref{5}) is different from that of the starting action (\ref{1}).
Indeed the equation (\ref{13}) leads to the zero value of the
energy of systems. To correct the situation and to get a
time-dependent Schr\"odinger equation for the parametrized system
(\ref{5}) we will proceed as follows.
 We regard the
following invariant action
\be
S_r = -\int p_t \left \{ -\frac {dt}{d\tau}(\tau) + N(\tau) \right \} d\tau.
\label{14}
\ee
Now $(t, p_t)$ is a pair of dynamic conjugated variables, $p_t$ is the
momentum corresponding to $t$. The action (\ref{14}) is invariant under
reparametrization (\ref{6}), if
\be
\delta p_t = a\dot p_t,\qquad\qquad \delta t = a\dot
t,\qquad\qquad \delta N = \frac{d}{d\tau}(aN), \label{15} \ee
since $\delta S_r = \int \frac{d}{d\tau}(a p_t N-ap_t \dot t) d
\tau$ is a total derivative. So, adding the action (\ref{14}) to
the action (\ref{5}) we obtain in the first order form the total
action $\tilde{\tilde{S}}= \tilde S + S_r$
\be
\tilde{\tilde{S}} = \int \left\{ p_i \dot x^i - NH_0(p, x) +
p_t(\dot t - N(\tau )) \right \} d\tau.
\label{16}
\ee
The action (\ref{16}) is invariant under the local transformation (\ref{6}),
if $N, x, p_t$ and $t$ transform according to (\ref{7}, \ref{15}).

So, we will proceed with the canonical quantization of the action
(\ref{14}-\ref{16}). Following the rules of this procedure we have two constraints
corresponding to the canonical variables $t$ and $p_t$
\be
\Pi_1 \equiv P_t - p_t = 0,  \qquad\qquad
\Pi_2 \equiv P_{p_t} = 0,
\label{17}
\ee
where $P_t = \frac{\partial \tilde{\tilde{L}}}{\partial {\dot t}} = p_t$ and
$P_{p_t} = \frac{\partial \tilde{\tilde{L}}}{\partial {\dot p_t}} = 0$ are the
momenta conjugated to $t$ and $p_t$, respectively.

The constraints (\ref{17}) are of the second class, and therefore,
they can be eliminated by the Dirac's procedure. Defining the matrix constraint
$C_{AB}$ with $(A, B= 1,2)$ as a Poisson bracket we find, that the only
non-zero matrix elements are
\be
C_{1,2} = \lbrace \Pi_1, \Pi_2 \rbrace =- 1, \qquad  C_{2,1} =
\lbrace \Pi_2, \Pi_1 \rbrace = 1,
\label{18}
\ee
with their inverse matrix elements $(C^{-1})^{1,2} = 1$ and $(C^{-1})^{2,1}
= -1$. The Dirac's brackets $\lbrace , \rbrace^{\ast}$ are defined by
\be
\lbrace A, B \rbrace^{\ast} = \lbrace A, B \rbrace - \lbrace A, \Pi_i\rbrace
(C^{-1})^{ij} \lbrace \Pi_j, B \rbrace.
\label{19}
\ee
The result of this procedure leads to the non-zero Dirac's brackets
relations
\be
\lbrace t, p_t \rbrace^{\ast} = 1.
\label{20}
\ee
Then, the canonical Hamiltonian obtained from the action
$\tilde{\tilde {S}}$ in (\ref{16}) has the form
\be
{\tilde{\tilde{H}}}_c = N(p_t + H_0),
\label{21}
\ee
and the total Hamiltonian is
\be
\tilde{\tilde {H_T}} = N(p_t + H_0) + u_N P_N,
\label{22}
\ee
where $u_N$ is the Lagrange multiplier associated to the constraint
$P_N =0$ in (\ref{10}), which must be conserved in the time, i.e.
\be
\dot P_N = \lbrace P_N, \tilde{\tilde{H_T}} \rbrace = -( p_t + H_0)
= 0,
\label{23}
\ee
which by definition is the first-class constraint. So, Hamiltonian's
equation of motion then yields
\be
\dot x_i = \lbrace x_i, \tilde{\tilde{H_T}} \rbrace = \frac{Np_i}{m},
\label{24}
\ee
\be
\dot p_i = \lbrace p_i, \tilde{\tilde{H_T}} \rbrace = -N\frac {dV}{dx_i},
\label{25}
\ee
\be
\dot N = \lbrace N, \tilde{\tilde{H_T}} \rbrace = u_N,
\label{26}
\ee
\be
\dot t = \lbrace t, \tilde{\tilde{H_T}} \rbrace = N,
\label{27}
\ee
\be
\dot p_t = \lbrace p_t, \tilde{\tilde{H_T}} \rbrace = 0.
\label{28}
\ee
The first two equations (\ref{24}) and (\ref{25}) are the equations of
motion for the physical degrees of freedom. The action (\ref{16}) contains
one extra canonical pair $(t, p_t)$ over (\ref{1}), but also contains the
constraint (\ref{23}). This constraint, being the only one, is of the
first-class. Furthemore, the action (\ref{16}) describes the same number of
independent
degrees of freedom as the action in (\ref{1}). The equation (\ref{26}) shows
that $N(\tau)$ is an arbitrary function playing the role of gauge field
of the reparametrization symmetry. If we take the gauge condition
$N(\tau)=1$, then as it follows from (\ref{27}), we have $t=\tau$. On the
level
of the equations of motion the action $S_r$ is zero, and inserting
 $N=\dot t$ in
the action $\tilde S$ in (\ref{5}), we can exclude the auxiliary
gauge field $N(\tau)$ and obtain Dirac's approach for
reparametrization invariant action in the case of non-relativistic
systems \cite{2,7,8}.

At the quantum level Dirac's brackets (\ref{20}) must be replaced by
the commutator
\be
[t, \hat p_t] =i\lbrace t, p_t \rbrace^{\ast} = i, \label{29} \ee
and the classical momentum $p_t$ by the operator $\hat p_t$ with
the representation $-i \frac {\partial}{\partial t}$ (we assume
units in which $\hbar=c=1$). Following the Dirac's canonical
quantization the first-class constraints must be imposed on the
wave function $\psi(x,t)$. So, the constraint (\ref{23}) may be
written as
\be
i \frac{d \psi(x^l, t)}{dt} = H_0 (-i \frac{d}{x^l}, x_m) \psi(x^l, t).
\label{30}
\ee
Hence, the inclusion in (\ref{5}) of an additional reparametrization
invariant action (\ref{14}) does not change the equations of motion
(\ref{24}, \ref{25}), but only the constraint (\ref{13}), which becomes
(\ref{23}). Thus, canonical quantization procedure applied to the
parametrized theory (\ref{16}) yields the correct equation for the
wave function $\psi$ (\ref{30}), which is just the conventional
time-dependent Schr\"odinger equation.

In the following two sections it will be shown, that the same procedure
without any difficulties can be extended to the relativistic and
supersymmetric cases.

\setcounter{equation}0\section{Relativistic Point Particle}

In this section we will consider a free relativistic particle. The
action in this case has the form
\be
S = - m\int \sqrt{1 - \dot x_i^2(t)}\,dt,
\label{31}
\ee
where $m, t$ and $x_i$ $(i = 1, 2, 3)$ are, respectively, the mass, proper
time and the position of the particle. After parametrization
$dt = N(\tau) d\tau$ the action (\ref{31}) becomes
\be
\tilde S = -m \int \sqrt{N^2(\tau) - \dot x_i^2 (\tau)}\, d\tau .
\label{32}
\ee
This action is invariant under the local time reparametrization
(\ref{6}), if $N(\tau)$ and $x_i(\tau)$ transform as (\ref{7}).
The canonical Hamiltonian in this case has the form
\be
\tilde H_c = NH_0 = N \left (\sqrt {p_i^2 + m^2} \right ),
\label{33}
\ee
where $p_i = \frac{\partial {\tilde L}}{\partial \dot x_i} =  \frac{m}{N}
\frac{\dot x_i}{\sqrt {1-\frac{\dot x_i}{N^2}}}$ is the canonical momentum
conjugate to dynamical variable $x_i$.

So, we will rewrite the action (\ref{32}) by considering (\ref{14}) in the
first order form, we get
\be
\tilde{\tilde{S}} = \int\left\{ p_i \dot x_i + p_0({\dot x}^0 - N) -
N\sqrt{p_i^2 + m^2}\right \} d\tau,
\label{34}
\ee
where we have $p_0 \equiv p_t$ and $x^0 \equiv t$. Following the analogous
procedure of the proceeding section, i.e. eliminating the second-class
constraints by means of Dirac's brackets (\ref{19}), we get the
relativistic canonical Hamiltonian
\be
\tilde{\tilde{H}}_c = NH = N \left ( \sqrt{p_i^2 + m^2} + p_0 \right ),
\label{35}
\ee
where $H$ is the classical relativistic constraint corresponding to the
action (\ref{34}). At the quantum level this constraint becomes condition
on the wave function $\psi$
\be
\left (-i \frac{d}{dx_0} + \sqrt{\hat p^2_i + m^2} \right )\psi (x_0, x_i)=0,
\label{36}
\ee
this is the time-dependent Schr\"odinger equation for the relativistic
free massive particle.

Note, that if we take the lapse function as
\be
N(\tau) = e(\tau)\frac{\sqrt{p^2_i + m^2}- p_0 }{2},
\label{37}
\ee
and putting it in (\ref{35}) we have then
\be
{\tilde{\tilde {H}}}_c = \frac{e(\tau)}{2} \left (\sqrt{p^2_i + m^2} -
p_0 \right) \left (\sqrt{p^2_i + m^2} + p_0 \right ) = \frac{e(\tau)}{2}
\left (p^2_i + m^2 - p^2_0 \right ).
\label{38}
\ee
Using the relations (\ref{37}), (\ref{7}) and (\ref{15}) for the
$N(\tau), p_i(\tau)$ and $p_0(\tau)$, it is easy to show, that $e(\tau)$
transforms as
\be
\delta e = (ae)^.,
\label{39}
\ee
corresponding to the transformation of $N(\tau)$ in (\ref{7}).

So, the action (\ref{34}) takes the form
\be
S= \int\left \{ p_\mu \dot x^\mu - e(\tau)\left(\frac{p^2_\mu +
m^2}{2}\right)\right \}
d\tau,
\label{40}
\ee
where $\mu = 0,1,2,3$. The action (\ref{40}) describes a massive
relativistic particle moving on the four dimensional space-time. The
$e(\tau)$ is an einbein, which plays the role of Lagrange multiplier.
Variation of the action (\ref{40}) with respect to $e(\tau)$ leads to the
relativistic constraint
\be
p^2_\mu + m^2 = 0,
\label{41}
\ee
which is nothing but the mass-shell condition. When we go over to quantum
mechanics, the constraint (\ref{41}) is replaced by the condition on the
scalar field $\phi$
\be
\left( \frac{\partial^2}{\partial x^2_0} - \frac {\partial^2}
{\partial x^2_i} + m^2 \right) \phi(x_0, x_i) = 0,
\label{42}
\ee
which is the Klein-Gordon equation.
Hence, inclusion of an additional action invariant under
reparametrization leads us to the Schr\"odinger time-dependent
equation for the wave function $\psi(x,t)$ in the case of
relativistic particle, and at the same time it leads to the
Klein-Gordon equation in the case of quantum scalar field
$\phi(t,x)$. In this approach it is not necessary to introduce
auxiliary time in order to obtain the Schr\"odinger equation
\cite{7}.

\setcounter{equation}0\section{n=2, d= 1 Supersymmetry}

In the global $n=2$ supersymmetric one dimensional quantum mechanics the
simplest action has the form \cite{10,13,14}
\be
S_{n=2} = \int \left \{\frac{\dot x^2}{2} - i\bar\chi \dot \chi - 2
\Big(\frac{\partial g}{\partial x} \Big)^2 - 2\frac{\partial^2g}{\partial x^2}
\bar\chi \chi \right\} dt,
\label{44}
\ee
where the overdote denotes derivatives with respect to $t$.
In the action (\ref{44}) $x$ is an even dynamical variable, unlike
$\chi$, which is odd. Note, that the action in (\ref{44}) is the
supersymmetric extension of (\ref{1}).

The corresponding supersymmetric Hamiltonian is
\be
H_0 = \frac{p^2}{2} + 2 \Big(\frac{\partial g}
{\partial x} \Big)^2 + 2 \frac{\partial^2 g}{\partial x^2}
\bar\chi \chi,
\label{45}
\ee
where $p = \dot x $, $\pi_{\chi} = -i\bar\chi$ and
$\pi_{\bar\chi}= -i\chi$ are the momenta conjugated to
$x, \chi$ and $\bar\chi$, respectively. The Dirac's brackets are
defined as
\be
\lbrace \chi, \bar\chi \rbrace^\ast = -i, \qquad\qquad
\lbrace x, p \rbrace^\ast = 1.
\label{46}
\ee
Applying the Noether theorema to the $n=2$ supersymmetry invariant action
one finds the corresponding conserved supercharges
\be
S= \left( ip + 2\frac{\partial g}{\partial x} \right) \chi, \qquad
\bar S = S^\dagger = \left( -ip + 2\frac{\partial g}{\partial x} \right)
\bar\chi,
\label{47}
\ee
and $F$, which is the generator of the $U(1)$ rotation on $\chi$
\be
F= \bar\chi \chi.
\label{48}
\ee
In terms of the Dirac's brackets (\ref{46}) the quantities $H_0$, $S$,
$\bar S$ and $F$ form a closed super-algebra
\begin{eqnarray}
\lbrace S, \bar S\rbrace^\ast &=& -2iH_0,\quad
\lbrace H_0, S\rbrace^\ast = \lbrace H_0, \bar S\rbrace^\ast = 0,
\quad \lbrace S, S\rbrace^\ast =
\lbrace \bar S, \bar S\rbrace^\ast = 0, \label{49} \\
\lbrace F, S\rbrace^\ast &=& iS, \quad
\lbrace F, \bar S\rbrace^\ast = -i\bar S. \nonumber
\end{eqnarray}

Now, our goal will be to obtain the time-dependent Schr\"odinger equation
for the supersymmetric case. The approach will be similar to that we have
followed earlier. Dirac's approach applied to the action (\ref{44}) for the
$n=2$ supersymmetric mechanics in the reparametrization invariant form
requires a modification. A direct way to construct such action is a
supersymmetric extension of the action (\ref{5}), including in the local
$n=2$ supersymmetry the transformations of reparametrization (\ref{6}). As
a consequence of this extension the new gauge fields $\psi(\tau),
\bar\psi(\tau)$ and $V(\tau)$ in the action will appear. These gauge
fields are the superpartners of the ``lapse function" $N(\tau)$.

In order to obtain the superfield formulation of the action (\ref{44}) the
transformation of the time reparametrization (\ref{6}) must be extended to the
$n=2$ local conformal time supersymmetry $(\tau, \theta, \bar\theta)$
\cite{15,16,17,18}. The transformations of the supertime $(\tau, \theta,
\bar\theta)$ can be written as

\begin{eqnarray}
\delta \tau &=& {I\!\!L}(\tau,\theta,\bar\theta) +
\frac{1}{2} \bar\theta D_{\bar\theta}{I\!\!L}(\tau,\theta,\bar\theta)
- \frac{1}{2} \theta D_{\theta}{I\!\!L}(\tau,\theta,\bar\theta),\nonumber\\
\delta \theta &=& \frac{i}{2} D_{\bar\theta} {I\!\!L}(\tau,\theta,\bar\theta),
\qquad
\delta \bar\theta = - \frac{i}{2} D_{\theta}{I\!\!L}
(\tau,\theta,\bar\theta),
\label{50}
\end{eqnarray}
with the superfunction ${I\!\!L}(\tau,\theta,\bar\theta)$ defined by

\be
{I\!\!L}(\tau,\theta,\bar\theta) = a(\tau) + i\theta \bar\beta^{\prime}(\tau)
+ i\bar\theta\beta^{\prime}(\tau) + b(\tau)\theta\bar\theta,
\label{51}
\ee
where $D_{\theta} = \frac{\partial}{\partial \theta} +
i\bar\theta \frac{\partial}{\partial \tau}$ and $ D_{\bar\theta} =
-\frac{\partial}{\partial \bar\theta} - i\theta\frac{\partial}{\partial\tau}$
are the supercovariant derivatives of the $n=2$ global supersymmetry,
$a(\tau)$ is a local time reparametrization parameter,
$\beta^{\prime}(\tau)$ is the Grassmann complex parameter of the $n=2$ local
conformal supersymmetry transformations and $b(\tau)$ is the
parameter of the local $U(1)$ rotations on the complex Grassmann
coordinates $\theta$ $(\bar\theta = \theta^{\dagger})$.

Then, the superfield generalization of the actions (\ref{5}) and (\ref{44}),
which are invariant under the $n=2$ local conformal supersymmetry
transformations (\ref{50}), has the form \cite{19,20}.
\be
\tilde S_{n=2} = \int \bigg\{ \frac{1}{2}{I\!\!N^{-1}} D_{\bar\theta}
\Phi D_{\theta}\Phi - 2g(\Phi) \bigg \} d\theta d\bar\theta d\tau,
\label{52}
\ee
where $g(\Phi)$ is the superpotential. The local supercovariant derivatives
have the form $\tilde D_{\theta} = {I\!\!N^{-\frac{1}{2}}} D_{\theta}$ and
$\tilde D_{\bar\theta} = {I\!\!N^{-\frac{1}{2}}}D_{\bar\theta}$. In the
superfield action (\ref{52}) ${I\!\!N}(\tau,\theta,\bar\theta)$ is absent
in the numerator of the second term, this is related to the fact that the
superjacobian of the transformations (\ref{50}), as well as the $BerE_B^A$,
is equal to one and the quantity $d\theta d\bar\theta d\tau$ is an invariant
volume.

In order to have the component action for (\ref{52}) we must
expand the superfields ${I\!\!N}$, $\Phi$ and the superpotential
$g(\Phi)$ in Taylor series  with respect to $\theta, \bar\theta$.

In the case of the real superfield ${I\!\!N}$ $(i.e. {I\!\!N}^{\dagger} =
{I\!\!N})$ we have the following expansion

\be
{I\!\!N}(\tau,\theta,\bar\theta) = N(\tau) + i\theta \bar\psi^{\prime}(\tau) +
i\bar\theta \psi^{\prime}(\tau) + V^{\prime}(\tau) \theta\bar\theta,
\label{53}
\ee
where $N(\tau)$ is the lapse function, $\psi^{\prime} = N^{1/2}(\tau)
\psi(\tau)$ and $V^{\prime}(\tau) = NV + \bar\psi \psi$. The components
$N, \psi, \bar\psi$ and $V$ of the superfield
${I\!\!N}(\tau,\theta,\bar\theta)$ are gauge fields of the one-dimensional
$n=2$ supergravity.

The superfield (\ref{53}) transforms as the one-dimensional vector field
under the local supersymmetric transformations (\ref{50})

\be
\delta {I\!\!N} = ({I\!\!L}{I\!\!N})^. + \frac{i}{2} D_{\bar\theta}
{I\!\!L} D_{\theta} {I\!\!N} + \frac{i}{2} D_{\theta}{I\!\!L}
D_{\bar\theta} {I\!\!N}.
\label{54}
\ee
The transformation law for the components $N(\tau), \psi(\tau),
\bar\psi(\tau)$ and $V(\tau)$ may be obtained from (\ref{54})

\begin{eqnarray}
\delta N &=& (aN)^. + \frac{i}{2}(\beta \bar\psi + \bar\beta \psi), \qquad
\delta V = (aV)^. + \dot{\hat b}, \label{55}\\
\delta \psi &=& (a\psi)^. + D\beta - \frac{i}{2} \hat b \psi, \qquad
\delta \bar\psi = (a\bar\psi)^. + D\bar\beta + \frac{i}{2} \hat b \bar\psi,
\nonumber
\end{eqnarray}
where $D\beta = \dot\beta + \frac{i}{2} V\beta$ and
$ D\bar\beta = \dot{\bar\beta} - \frac{i}{2} V\beta$ are the $U(1)$
covariant derivatives and $\hat b = b - \frac{1}{2N}(\beta \bar\psi -
\bar\beta \psi)$.

For the real scalar matter superfield $\Phi(\tau,\theta,\bar\theta)$
we have
\be
\Phi(\tau,\theta,\bar\theta) = x(\tau) + i\theta\bar\chi^{\prime}(\tau) +
i\bar\theta \chi^{\prime}(\tau) + F^{\prime}(\tau) \theta\bar\theta,
\label{56}
\ee
where $\chi^{\prime} = N^{1/2} \chi(\tau)$ and $F^{\prime} = NF +
\frac{1}{2}(\bar\psi\chi - \psi\bar\chi)$.

The transformations law for the superfield $\Phi(\tau,\theta,\bar\theta)$
is
\be
\delta \Phi= {I\!\!L} \dot\Phi + \frac{i}{2}D_{\bar\theta}{I\!\!L}
D_{\theta} \Phi + \frac{i}{2} D_{\theta} {I\!\!L} D_{\bar\theta} \Phi.
\label{57}
\ee
The component $F(\tau)$ in (\ref{56}) is an auxiliary degree of freedom
(non-dynamical variable), $\chi(\tau)$ and $\bar\chi(\tau)$ are the
``fermionic" superpartners of the $x(\tau)$. Their transformations law
has the form
\begin{eqnarray}
\delta x &=& a \dot x + \frac{i}{2}(\beta \bar\chi + \bar\beta \chi), \qquad
\delta F = a \dot F + \frac{1}{2N} (\bar\beta \tilde D \chi -
\beta \tilde D \bar\chi), \label{58}\\
\delta \chi &=& a \dot \chi + \frac{\beta}{2}\left (\frac{D \chi}{N} +
iF \right ) - \frac{i}{2} \hat b \chi, \qquad
\delta \bar\chi = a\dot {\bar\chi} + \frac{\bar\beta}{2}
\left (\frac{D \chi}{N} - iF \right ) + \frac{i}{2}\hat b \bar\chi,
\nonumber
\end{eqnarray}
where $Dx = \dot x - \frac{i}{2}(\psi \bar\chi + \bar\psi \chi)$,
$\tilde D\chi = D\chi - \frac{i}{2}(\frac{D\chi}{N} + iF)$ are the
supercovariant derivatives and $D\chi = \dot\chi + \frac{i}{2} V\chi$.

It is clear, that the superfield action (\ref{52}) is invariant under the
$n=2$ local conformal time supersymmetry. Now, we can write the expression
under the integral (\ref{52}) by means of certain superfunction
$f({I\!\!N}, \Phi)$. Then, the infinitesimal small transformations of the
action (\ref{52}) under the superfield transformations (\ref{54},\ref{57})
have the form
\be
\delta \tilde S_{n=2}= \frac{i}{2} \int \left\{
D_{\bar\theta}({I\!\!L} D_{\theta} f) +
D_{\theta}({I\!\!L}D_{\bar\theta} f) \right\} d\theta d\bar\theta
d\tau. \label{59} \ee We can see, that the integrand is a total
derivative, $i.e.$ the action (\ref{52}) is invariant under the
$n=2$ local conformal time supersymmetry.

After integration over the Grassmann complex coordinates $\theta$
and $\bar\theta$ we find the component action, where $F(\tau)$ is
an auxiliary field, and it can be eliminated using its equation of
motion. Finally, the action $\tilde S_{n=2}$, in terms of the
components of the superfields ${I\!\!N}$ and $\Phi$, takes the
form
\be
\tilde S_{n=2} = \int \left\{ \frac{(Dx)^2}{2N} - i\bar\chi D\chi -
2N \left (\frac{\partial g}{\partial x}\right)^2 - 2N \frac{\partial^2 g}
{\partial x^2} \bar\chi \chi +
\frac{\partial g}{\partial x}(\bar\psi \chi - \psi \bar\chi) \right\}d\tau,
\label{60}
\ee
where $Dx$ and $D\chi$ are defined above.

The action (\ref{60}) does not contain the kinetic terms for $N,
\psi, \bar\psi$ and $V$, they are not dynamical. This fact is
reflected in the primary constraints
\begin{eqnarray}
P_N &=& \frac{\partial \tilde L_{n=2}}{\partial \dot N} = 0, \qquad
P_\psi = \frac{\partial \tilde L_{n=2}}{\partial \dot \psi} = 0, \qquad
P_{\bar\psi} = \frac{\partial \tilde L_{n=2}}{\partial \dot {\bar\psi}} = 0,
\label{61}\\
P_V  &=& \frac{\partial \tilde L_{n=2}}{\partial \dot V} = 0,
\nonumber
\end{eqnarray}
where $P_N, P_\psi, P_{\bar\psi}$ and $P_V$ are the canonical momenta
conjugated to $N, \psi, \bar\psi$ and $V$, respectively.

Then, the canonical Hamiltonian for the action $\tilde S_{n=2}$ in (\ref{60})
can be calculated in the usual way
\be
\tilde H_c = NH_0 + \frac{\bar \psi}{2} S - \frac{\psi}{2} \bar S +
\frac{V}{2} F,
\label{62}
\ee
where $H_0, S,\bar S$ and $F$ are defined in (\ref{45}, \ref{47}, \ref{48}).
Therefore, the total Hamiltonian is
\be
\tilde H_T = \tilde H_c + u_N P_N + u_\psi P_\psi + u_{\bar\psi}
P_{\bar\psi} + u_V P_V.
\label{63}
\ee
The secondary constraints are first-class constraints
\be
H_0 = 0, \qquad S=0, \qquad \bar S = 0, \qquad F=0,
\label{64}
\ee
which are obtained using the standard Dirac's procedure, $i.e.$, the time
derivatives of the primary constraints must be vanishing for all the $p$,
$x$, $\pi_{\chi}$, $\pi_{\bar\chi}$, $\chi$ and $\bar\chi$,
that satisfy the equation of motion.

In the quantum theory the first-class constraints (\ref{64})
associated with the invariance of the action (\ref{60}) become
conditions on the wave function $\psi = \psi(x, \chi,\bar\chi)$.
The quantum constraints are
\begin{eqnarray}
H_0 \psi &=& 0,\quad
S \psi = \bar S \psi = 0, \quad F\psi = 0,
\label{65}
\end{eqnarray}
which are obtained when we change the classical dynamical variables by their
corresponding operators. The first equation in (\ref{65}) is the Schr\"odinger
equation, a state with zero energy. Therefore, we have the time-independent
Schr\"odinger equation, this fact is due to the invariance under the
reparametrization symmetry of the action (\ref{60}), this problem is
well-known as the ``problem of time" \cite{1,2,3,4,5,6}.

So, in order to have a time-dependent Schr\"odinger equation for the
supersymmetric quantum mechanics, we consider the generalization of the
reparametrization invariant action $S_r$ in (\ref{14}). In the case of $n=2$
local supersymmetry it has the superfield form
\begin{eqnarray}
S_{r(n=2)} &=&- \int \left \{ {I\!\!P} - \frac{i}{2}{I\!\!N^{-1}}
(D_{\bar\theta}{\bf T} D_{\theta} {I\!\!P} - D_{\bar\theta}{I\!\!P}
D_{\theta}{\bf T}) \right \} d\theta d\bar\theta d\tau.
\label{66}
\end{eqnarray}

The action (\ref{66}) is determined in terms of the new superfields
${\bf T}$ and ${I\!\!P}$. The superfield ${\bf T}$ is determined by
the odd complex time $\eta(\tau)$ and $\bar\eta(\tau)$, which are the
superpartners of the time $t(\tau)$ and one auxiliary field
$m^{\prime}(\tau)$. Explicitly, we have
\be
{\bf T}(\tau, \theta, \bar\theta) = t(\tau) + \theta \eta^{\prime}(\tau) -
\bar\theta \bar\eta^{\prime}(\tau) + m^{\prime}(\tau)\theta \bar\theta,
\label{67}
\ee
where $\eta^{\prime}(\tau) = N^{1/2}(\tau) \eta(\tau)$ and
$m^{\prime}(\tau) = Nm + \frac{i}{2}(\bar\psi \bar\eta + \psi\eta)$. The
transformation rule for the superfield ${\bf T}(\tau,\theta,\bar\theta)$
under the $n=2$ local conformal supersymmetry transformations (\ref{50}) is
\be
\delta {\bf T} = {I\!\!L} \dot{\bf T} + \frac{i}{2} D_{\bar\theta}
{I\!\!L} D_{\theta}{\bf T} + \frac{i}{2} D_{\theta} {I\!\!L}
D_{\bar\theta} {\bf T}.
\label{68}
\ee

The superfield ${I\!\!P}(\tau,\theta,\bar\theta)$ has the form
\be
{I\!\!P}(\tau. \theta, \bar\theta) = \rho(\tau) +
i\theta p^{\prime}_{\bar\eta}(\tau) + i\bar\theta p^{\prime}_{\eta}(\tau) +
p^{\prime}_t(\tau) \theta\bar\theta,
\label{69}
\ee
where $p^{\prime}_\eta(\tau) = N^{1/2}p_{\eta}(\tau)$ and
$p^{\prime}_t = Np_t + \frac{1}{2}(\bar\psi p_{\eta} - \psi p_{\bar\eta})$.
$p_{\eta}$ and $p_{\bar\eta}$ are the odd complex momenta, $i.e.$
superpartners of the momentum $p_t$. The superfield ${I\!\!P}$ transforms as
\be
\delta {I\!\!P} = {I\!\!L} \dot {I\!\!P} + \frac{i}{2}D_{\bar\theta}
{I\!\!L} D_{\theta}{I\!\!P} + \frac{i}{2} D_{\theta}{I\!\!L}
D_{\bar\theta}{I\!\!P}.
\label{70}
\ee
It is easy to show, that the infinitesimal small transformations of the
action $S_{r(n=2)}$ under the transformations (\ref{54}, \ref{68},
\ref{70}) is a total derivative, then the action $S_{r (n=2)}$ is invariant
under the $n=2$ local supersymmetric transformations (\ref{50}).

After integration over $\theta$ and $\bar\theta$ the action (\ref{66}) may
be written in its component form. We obtain
\begin{eqnarray}
S_{r(n=2)} &=&- \int \left\{ p_t(N - \dot t) + i\dot\eta p_{\eta} +
i\dot{\bar\eta} p_{\bar\eta} + \frac{\bar \psi}{2}(p_{\eta} - \bar\eta p_t)
-\frac{\psi}{2} (p_{\bar\eta} - \eta p_t) \right.\label{71}\\
&+& \left. \frac{V}{2} (\eta p_{\eta} - \bar\eta p_{\bar\eta}) +
m \dot \rho - \frac{i}{2}m \psi p_{\bar\eta} - \frac{i}{2}m \bar\psi p_{\eta}
\right\} d\tau. \nonumber
\end{eqnarray}
One can show that the variables $\rho$ and $m$ are auxiliary, in
the sense, that they can be eliminated from the physical variables
by some unitary transformation.

So, the component action has the final form
\begin{eqnarray}
S_{r(n=2)} &=&- \int \left\{ p_t(N - \dot t) + i\dot\eta p_{\eta}
+ i\dot{\bar\eta} p_{\bar\eta} + \frac{\bar\psi}{2}(p_{\eta} - \bar\eta p_t) -
\frac{\psi}{2}(p_{\bar\eta} - \eta p_t) \right.\label{73}\\
&+& \left. \frac{V}{2}(\eta p_{\eta} - \bar\eta p_{\bar\eta}) \right\} d\tau.
\nonumber
\end{eqnarray}
We can see from (\ref{73}) that the action $S_{r (n=2)}$ contains
$S_r$ term and the additional terms related to the $n=2$ local supersymmetry
transformations of the components of the superfields $\bf T$,
${I\!\!P}$ and ${I\!\!N}$.

Varying the action (\ref{73}) with respect to $p_t, p_\eta$ and
$p_{\bar\eta}$ we obtain the relations between $N, \psi, \bar\psi, t, \eta$
and $\bar\eta$, which are the generalization of (\ref{27})
\be
N(\tau) = \dot t + \frac{1}{2}\bar\psi(\tau) \bar\eta(\tau) -
\frac{1}{2}\psi(\tau) \eta(\tau), \qquad \psi = 2iD\bar\eta, \qquad
\bar\psi = -2iD\eta,
\label{74a}
\ee
where $D\eta = \dot \eta - \frac{i}{2}V\eta$ and
$D\bar\eta = \dot {\bar\eta} + \frac{i}{2}V\bar\eta$ are the $U(1)$
supercovariant derivatives. Fulfilling the relations (\ref{74a}) the action
(\ref{73}) vanishes.

Proceeding to the Hamiltonization, in addition to the second-class
constraints obtained in (\ref{17}) corresponding to the canonical
variables $t$ and $p_t$, we have the following constraints
\begin{eqnarray}
\Pi_3(\eta) &=& P_\eta + ip_\eta = 0, \qquad \Pi_4(p_\eta) =
P_{p_\eta} = 0, \label{74} \\
\Pi_5(\bar\eta) &=& P_{\bar\eta} + ip_{\bar\eta} = 0, \qquad
\Pi_6 (p_{\bar \eta}) = P_{p_{\bar \eta}} = 0, \nonumber
\end{eqnarray}
where $P_\eta = \frac{\partial L_{r(n=2)}}{\partial {\dot \eta}}$,
$P_{p_\eta} = \frac{\partial L_{r(n=2)}}{\partial {\dot p_\eta}}$ are the odd
momenta conjugated to $\eta$, $p_\eta$ and their respective complex
conjugate. We define the odd canonical Poisson brackets as
\be
\lbrace \eta, P_\eta \rbrace = 1, \qquad \lbrace p_\eta, P_{p_\eta} \rbrace
=1.
\label{75}
\ee
So, the constraints (\ref{74}) are of the second-class. Defining the matrix
(symmetric for the Grassmann variables) constraint
$C_{ik}$ $(i,k = \eta, p_\eta, \bar\eta, p_{\bar\eta})$ as the odd Poisson
brackets, we have the following non-zero matrix elements
\be
C_{\eta, p_{\eta}} = C_{p_{\eta}, \eta} = \lbrace \Pi_3, \Pi_4 \rbrace = i,
\qquad C_{{\bar\eta}, p_{\bar\eta}} = C_{p_{\bar\eta}, {\bar\eta}} =
\lbrace \Pi_5, \Pi_6 \rbrace = i
\label{76}
\ee
with their inverse matrix $(C^{-1})^{\eta, p_{\eta}} =- i$ and
$(C^{-1})^{\bar\eta, p_{\bar\eta}} =- i$. Using the Dirac's brackets
$\lbrace , \rbrace^\ast$ defined in (\ref{19}) we obtain, that the
only non-zero matrix elements are
\be
\lbrace \eta, p_\eta \rbrace^{\ast} = i, \qquad
\lbrace \bar\eta, p_{\bar\eta} \rbrace^{\ast} = i.
\label{77}
\ee
So, if we take the additional term (\ref{71}) the full action will be
\be
\tilde {\tilde{S}} = \tilde S_{n=2} + S_{r (n=2)}.
\label{78}
\ee
Then, the canonical Hamiltonian for the action $\tilde {\tilde {S}}$
will have the form
\be
{\tilde {\tilde {H}}}_c = N(p_t + H_0) - \frac{\psi}{2}(S_{\bar\eta} +
\bar S) + \frac{{\bar \psi}}{2}(-S_{\eta} + S) + \frac{V}{2}(F_\eta + F),
\label{79}
\ee
where $S_{\eta} = (-p_{\eta} + \bar\eta p_t)$,
$S_{\bar\eta} = (p_{\bar\eta} - \eta p_t)$ and
$F_{\eta} = (\eta p_{\eta} - \bar\eta p_{\bar\eta})$.

Then the total Hamiltonian may be written as
\be
{\tilde {\tilde {H}}}_T = {\tilde {\tilde {H}}}_c + u_N P_N +
u_{\psi} P_{\psi} + u_{\bar\psi} P_{\bar\psi} + u_V P_V.
\label{80}
\ee
Due to the conditions
\be
\dot P_N = \dot P_\psi = \dot P_{\bar\psi} = \dot P_V = 0,
\label{81}
\ee
we now have the first-class constraints
\be
H = p_t + H_0, \qquad Q_{\eta} = -S_{\eta} + S, \qquad
Q_{\bar\eta} =  S_{\bar\eta} + \bar S, \qquad {\cal F} = F_{\eta} + F.
\label{82}
\ee
The constraints (\ref{82}) form a closed superalgebra with respect to the
Dirac's brackets
\begin{eqnarray}
\lbrace Q_{\eta}, Q_{\bar\eta} \rbrace^{\ast} &=& -2iH, \quad
\lbrace H, Q_{\eta} \rbrace^{\ast} =
\lbrace H, Q_{\bar\eta} \rbrace^{\ast} = 0, \label{83} \\
\lbrace {\cal F}, Q_{\eta} \rbrace^{\ast} &=& iQ_{\eta}, \quad
\lbrace {\cal F}, Q_{\bar\eta} \rbrace^{\ast} = -iQ_{\bar\eta}. \nonumber
\end{eqnarray}
After quantization the Dirac's brackets (\ref{77}) become anticomutator
for the odd variables
\begin{eqnarray}
\lbrace \eta, p_{\eta} \rbrace &=& i\lbrace \eta, p_{\eta}\rbrace^{\ast}
=- 1, \qquad \lbrace \bar\eta, p_{\bar\eta} \rbrace =
i \lbrace \bar\eta, p_{\bar\eta} \rbrace^{\ast}=-1,
\label{84}
\end{eqnarray}
with the operator representation $p_{\eta} =-
\frac{\partial}{\partial \eta}$ and $p_{\bar\eta} =-
\frac{\partial}{\partial {\bar\eta}}$.
In order to obtain the quantum expression for $H, Q_{\eta}, Q_{\bar\eta}$
and $\cal F$  we use the operator representation $p = -i\frac{d}{dx}$ and
$\chi$, $\bar\chi$ as $\lbrace \chi, \bar\chi \rbrace = 1$,
$\chi = \sigma_{(-)}$ and $\bar\chi = \sigma_{(+)}$, where
$\sigma_{\pm} = \frac{1}{2} (\sigma_1 \pm i\sigma_2)$ in our case
for the generators (\ref{84}) on the quantum level, we have
\begin{eqnarray}
H &=& -i\frac{d}{dt} + H_0(p,x,\chi,\bar\chi),\qquad
Q_{\eta} =- \left(\frac{\partial}{\partial \eta} - i\bar\eta
\frac{\partial}{\partial t} \right) + S(p, x, \chi), \label{85} \\
Q_{\bar\eta} &=& \left(-\frac{\partial}{\partial {\bar\eta}} +
i\eta \frac{\partial}{\partial t}\right) + \bar S(p, x, \bar\chi), \qquad
{\cal F}= \left(\bar\eta \frac{\partial}{\partial {\bar\eta}} -
\eta\frac{\partial}{\partial \eta}\right) + F(\chi, \bar\chi), \nonumber
\end{eqnarray}
where $ H_0 = - \frac{d^2}{dx^2} + 2(\frac{\partial g}
{\partial x})^2 + \frac{d^2 g}{\partial x^2} [\bar\chi, \chi]$
and $F= \frac{1}{2} [\bar\chi, \chi] = \frac{1}{2} \sigma_3$. In
(\ref{85}) $ S_{\eta} = \frac{\partial}{\partial \eta} - i\bar\eta
\frac{\partial}{\partial t}$ and $S_{\bar\eta} = - \frac{\partial}
{\partial {\bar\eta}} + i\eta \frac{\partial}{\partial t}$ are the
generators of supertranslations on the superspace with coordinates
$(t, \eta, \bar\eta)$ and $p_t = -i\frac{\partial}{\partial t}$ is
the ordinary time translation operator
\be
\lbrace S_{\eta}, S_{\bar\eta} \rbrace = 2i \frac{\partial}{\partial t},
\label{86}
\ee
and $F_{\eta} = -\eta \frac{\partial}{\partial \eta} +
\bar\eta \frac{\partial}{\partial {\bar\eta}}$ is the generator of the
$U(1)$ rotation on the complex Grassmann coordinates $\eta$ $(\bar\eta =
\eta^\dagger)$. The algebra of the quantum generators $H, S, \bar S$ and
$ F$ is a closed superalgebra
\begin{eqnarray}
\lbrace S, \bar S\rbrace &=& 2H_0, \qquad
\lbrack S, H_0 \rbrack = \lbrack \bar S, H_0 \rbrack =
\lbrack F, H_0 \rbrack = 0, \label{87} \\
\lbrack F, S \rbrack &=& - S, \qquad
\lbrack F, \bar S \rbrack = \bar S, \qquad
S^2 = {\bar S}^2 = 0, \nonumber
\end{eqnarray}
the conserved quantities are $H, S, \bar S$ and $F$. We can see, that the
generators $H, Q_{\eta}, Q_{\bar \eta} $ and
$\cal F$ satisfy the same superalgebra
\begin{eqnarray}
\lbrace Q_{\eta}, Q_{\bar\eta} \rbrace &=& 2H, \qquad
\lbrack Q_{\eta}, H \rbrack = \lbrack Q_{\bar\eta}, H \rbrack =
\lbrack {\cal F}, H \rbrack = 0, \label{88}\\
\lbrack {\cal F}, Q_{\eta} \rbrack &=& - Q_{\eta}, \qquad
\lbrack {\cal F}, Q_{\bar\eta} \rbrack = Q_{\bar\eta}, \qquad
Q_{\eta}^2 = Q_{\bar\eta}^2 = 0 \nonumber.
\end{eqnarray}
In the quantum theory the first-class constraints (\ref{85}) become
conditions on the wave function $\Psi$. So, we have the supersymmetric
quantum constraints
\begin{eqnarray}
H\Psi&=&0, \quad Q_\eta \Psi= 0, \quad
Q_{\bar\eta} \Psi = 0, \quad {\cal F}\Psi=0.
\label{89}
\end{eqnarray}

We will search the wave function in the superfield form, we regard
\begin{eqnarray}
\Psi(t, \eta, \bar\eta, \chi, \bar\chi) &=& \psi(t, x, \chi,
\bar\chi) + i\eta \sigma(t, x, \chi, \bar\chi) +
i\bar\eta \phi(t, x, \chi, \bar\chi) + \label{90}\\
&& + \zeta(t, x, \chi, \bar\chi) \eta \bar\eta \nonumber.
\end{eqnarray}
This wave function must satisfy the quantum constraints (\ref{89}). In
(\ref{90}) $\psi, \zeta$ are even components of the wave function, unlike
$\sigma, \phi$, which are odd. We take the constraints
\begin{eqnarray}
Q_{\eta} \Psi &=& 0, \qquad Q_{\bar\eta} \Psi = 0.
\label{91}
\end{eqnarray}
Due to the algebra (\ref{88}) we have
\be
\lbrace Q_{\eta}, Q_{\bar\eta} \rbrace \Psi = 2 H \Psi = 0.
\label{92}
\ee
This is the time-dependent Schr\"odinger equation for the supersymmetric
quantum mechanics.

The condition (\ref{91}) leads to the following form of the wave
function
\be
\Psi_{\ast} = \psi + \eta (S\psi) + \bar\eta (\bar S \psi) -
\frac{1}{2} (\bar S S - S \bar S) \psi \eta \bar\eta,
\label{93}
\ee
then, $Q_{\eta} \Psi$ has the form
\begin{eqnarray}
Q_{\eta} \Psi_{\ast} &=& \bar\eta (i \frac{d\psi}{dt} -\frac{1}{2}
\lbrace S, \bar S \rbrace \psi ) \label{94} \\
&+& \eta \bar\eta S(i \frac{d \psi}{dt}
- \frac{1}{2} \lbrace S, \bar S \rbrace \psi) =0 \nonumber,
\end{eqnarray}
this is the standard time-dependent Schr\"odinger equation
\be
i \frac{d \psi(t,\chi,\bar\chi)}{dt} = H_0 (p, x, \chi, \bar\chi)
\psi(t, x, \chi,\bar\chi),
\label{95}
\ee
due to the relation $H_0 = \frac{1}{2}\lbrace S, \bar S\rbrace$. If we put
in the Schr\"odinger equation (\ref{95}) the condition of
the stationary states given by $\frac{d\psi}{dt} = 0$, we will have
$H_0 \psi = 0$ and due to the algebra (\ref{87}) we obtain
$ S \psi = \bar S \psi = 0$ and the wave function $\Psi_{\ast}$
becomes wave function $\psi (x, \chi, \bar\chi)$ \cite{10,11,19,20}.

\setcounter{equation}0\section{Conclusions}

Without any difficulties our procedure may be generalized to D-dimensional
extended supersymmetry mechanics \cite{14,21}. This is due to the
fact, that the full algebra of the transformations is closed on off-shell,
and it is a $n=2$ local conformal supersymmetry. So, our procedure
represents a direct possibility to apply the Batalin-Vilkovsky formalism
\cite{22,23,24} to supersymmetric systems.

In this work we have considered systems (including susy), which are not
para\-me\-trized. Such systems always may be done in a parametrized invariant
form. For this purpose we must include auxiliary gauge degree of freedom.
Hence, the constraint system contains generator of reparametrization, which
is the Hamiltonian generator. Its operator must annihilate the physical
states, this leads to time-independet Schr\"odinger equation $H_0\psi = 0$
for states with zero energy.

In order to have a time-dependent Schr\"odinger equation, $i.e.$ to describe
the quantum evolution of a system, as we shown in this work, an
additional invariant action $S_r$ may be always constructed. The additional
action does not change the equation of motions, but the constraint system,
which becomes time-dependent Schr\"odinger equation. From our point of view,
this fact is very important in those cases, when starting systems
are invariant under reparametrization of time, such systems as: general
relativity, cosmological models, string theories. These theories contain
auxiliary additional gauge degree of freedom (lapse and shift functions)
\cite{25}. Such theories have the problem which in literature is known as the
``problem of time" \cite{1,3}. For instance, the Wheeler-DeWitt equation
\cite{26}.

Naturally, the question arising as a result of this work is: could we
construct an additional invariant under general covariant transformations
action? If the result of this question is positive, then the additional
action will remain without any changes the equations for the
physical degree of freedom of the system, but the constraint will be
modified leading to time-dependent Schr\"odinger equation.

\vspace{.5cm}

\noindent {\bf Acknowledgments.}
We are grateful to A. Ivanov, S. Krivonos, J.L. Lucio, J.A. Nieto,
I. Lyanzuridi, L. Marsheva, O. Obreg\'on J. Socorro and M. Tsulaia for their
interest in the work and useful comments. This research was supported in
part by CONACyT under the grant 28454E. Work of A.P. was supported in
part by INTAS Grant 96-0538 and by the Russian Foundation of Basic Research,
Grant 99-02-18417. One of us (J.J. Rosales) would like to thank CONACyT for
support under Estancias Posdoctorales en el Extranjero and Instituto de
F\'{\i}sica de la Universidad de Gto. for its hospitality during the final
stages of this work.


\newpage
\begin{center} {\large \bf REFERENCES}
\end{center}
\begin{thebibliography}{99}
\bibitem{1}  K. Kucha$\hat {r}$, in {\it Quantum Gravity 2}, edited by C.J.
             Isham, R. Penrose, and D.W. Sciama (Clarendon Press, Oxford,
             1981), p.329.

\bibitem{2}  P.A.M. Dirac, {\it Lectures on Quantum Mechanics} (Yeshiva Univ.
             Press, New York, 1967).

\bibitem{3}  M. Henneaux and C. Teitelboim, {\it Quantization of Gauge
             Systems} (Princeton Univ. Press, Princeton, NJ. 1992).

\bibitem{4}  C.J. Isham and K. Kucha$\hat {r}$, Ann. Phys. (N.Y.), {\bf 164},
             316 (1985).

\bibitem{5}  D.M. Gitman and I.V. Tyutin, {\it Quantization of Fields with
             Constraints} (Springer-Verlag Berlin Heidelberg, 1990).


\bibitem{6}  G. F\"ulop, D.M. Gitman and I.V. Tyutin, Int. Jour. of
             Theor. Phys. {\bf 38}, 1941 (1999).

\bibitem{7}  J.L. Lucio M., J. Antonio Nieto and J. David Vergara, Phys.
             Lett. {\bf A 219}, 150 (1996).

\bibitem{8}  J.B. Hartle, Class. and Quantum Grav. {\bf 13}, 361 (1996).

\bibitem{9}  E. Witten, Nucl. Phys. {\bf B 185},513 (1981).

\bibitem{10}  F. Coopen, A. Khare and U. Sukhatme, Phys. Rep. {\bf 251},
             267 (1995).

\bibitem{11} P. Salomonson and M.B. Halpern, Nucl. Phys. {\bf B 196},
             509 (1982).

\bibitem{12} V.I. Tkach, J.J. Rosales and O. Obregon, Class. and Quantum
             Grav. {\bf 13}, 2349 (1996); C. Teitelboim, Phys. Rev.
             {\bf D 25}, 3159 (1982).

\bibitem{13} C.V. Suluman, J. Phys. {\bf A 18}, 2917 (1985).

\bibitem{14} V.P. Berezevoj and A.I. Pashnev, Class. and Quantum Grav.
             {\bf 8}, 484 (1991).

\bibitem{15} L. Brink, P. DiVecchia and P. Howe, Nucl. Phys. {\bf B 118},
             76 (1977).

\bibitem{16} V.D. Gershun and V.I. Tkach, JETP Lett. {\bf 29}, 288 (1979).

\bibitem{17} P. Howe, S. Penati, M. Pernici and P. Townsend, Phys. Lett.
             {\bf B 215}, 555 (1988).

\bibitem{18} D.P. Sorokin, V.I. Tkach and D.V. Volkov, Mod. Phys. Lett.
             {\bf A 4}, 901 (1989).

\bibitem{19} V.I. Tkach, J.J. Rosales and J. Socorro, Class. and Quantum Grav.
             {\bf 16}, 797 (1999).

\bibitem{20} O. Obreg\'on, J.J. Rosales, J. Socorro and V.I. Tkach, Class.
             and Quantum Grav. {\bf 16}, 2861 (1999).

\bibitem{21} E.A. Ivanov, S.O. Krivonos and A.I. Pashnev, Class. and
             Quantum Grav. {\bf 8}, 19 (1991).

\bibitem{22} I.A. Batalin and G.A. Vilkovsky, Phys. Lett. {\bf B 102},
             27 (1981).

\bibitem{23} E.S. Fradkin and G.A. Vilkovsky, Phys. Lett. {\bf B 55},
             224 (1975).

\bibitem{24} I.A. Batalin and G.A. Vilkovsky, Phys. Lett. {\bf B 69},
             309 (1977).

\bibitem{25} W. Misner, K.S. Thorne and J.A. Wheeler, {\rm Gravitation}
             (W.M. Freeman, San Francisco, 1970).

\bibitem{26} B.S. De Witt, Phys. Rev. {\bf D 160}, 1113 (1967).


\end{thebibliography}
\end{document}